\documentclass[12pt]{article}
\usepackage{amsfonts}
\usepackage{latexsym}
\usepackage{times}

\catcode`\@=11

\newcount\hour
\newcount\minute
\newtoks\amorpm \hour=\time\divide\hour by 60\minute
=\time{\multiply\hour by 60 \global\advance\minute by-\hour}
\edef\standardtime{{\ifnum\hour<12 \global\amorpm={am}%
        \else\global\amorpm={pm}\advance\hour by-12 \fi
        \ifnum\hour=0 \hour=12 \fi
        \number\hour:\ifnum\minute<10
        0\fi\number\minute\the\amorpm}}
\edef\militarytime{\number\hour:\ifnum\minute<10
0\fi\number\minute}

\def\draftlabel#1{{\@bsphack\if@filesw {\let\thepage\relax
   \xdef\@gtempa{\write\@auxout{\string
      \newlabel{#1}{{\@currentlabel}{\thepage}}}}}\@gtempa
   \if@nobreak \ifvmode\nobreak\fi\fi\fi\@esphack}
        \gdef\@eqnlabel{#1}}
\def\@eqnlabel{}
\def\@vacuum{}
\def\marginnote#1{}
\def\draftmarginnote#1{\marginpar{\raggedright\scriptsize\tt#1}}
\def\draft{
        \pagestyle{plain}
        \overfullrule=2pt
        \oddsidemargin -.5truein
        \def\@oddhead{\sl \phantom{\today\quad\militarytime} \hfil
        \smash{\Large\sl DRAFT} \hfil \today\quad\militarytime}
        \let\@evenhead\@oddhead
        \let\label=\draftlabel
        \let\marginnote=\draftmarginnote
        \def\ps@empty{\let\@mkboth\@gobbletwo
        \def\@oddfoot{\hfil \smash{\Large\sl DRAFT} \hfil}
        \let\@evenfoot\@oddhead}
        \def\@eqnnum{(\theequation)\rlap{\kern\marginparsep\tt\@eqnlabel}%
        \global\let\@eqnlabel\@vacuum}  }

\newcommand{\rf}[1]{(\ref{#1})}
\renewcommand{\theequation}{\thesection.\arabic{equation}}
\renewcommand{\thefootnote}{\fnsymbol{footnote}}
\newcommand{\newsection}{    
\setcounter{equation}{0}\section}
\def\appendix#1{\addtocounter{section}{1}\setcounter{equation}{0}
\renewcommand{\thesection}{\Alph{section}}
\section*{Appendix \thesection\protect\indent \parbox[t]{11.15cm}{#1}}
\addcontentsline{toc}{section}{Appendix \thesection\ \ \ #1}}

\def\be{\begin{equation}}
\def\ee{\end{equation}}
\def\beq{\begin{eqnarray}}
\def\eeq{\end{eqnarray}}

\def\psik{|\psi\rangle}
\def\psibr{\langle\psi|}

\def\epsilonk{|\epsilon\rangle}

\def\parline{\,\partial\kern -0.55em /\,\,}

\def\FF{{\cal F}}
\def\LL{{\cal L}}
\def\MM{{\cal M}}

\def\smD{{\scriptscriptstyle D}}

\def\Dline{{D \kern-0.6em  /  }\,\, }
\def\parline{\,\partial\kern -0.55em /\,\,}

\def\fwt{\widetilde{f}}
\def\mwt{\widetilde{m}}

\def\smzero{{\scriptscriptstyle (0)}}
\def\smone{{\scriptscriptstyle (1)}}
\def\smtwo{{\scriptscriptstyle (2)}}

\def\mas{{\rm m}}
\def\coscon{\rho}

\begin{document}


\begin{flushright}
FIAN/TD/09-06 \\
hep-th/0609029
\end{flushright}

\vspace{1cm}

\begin{center}

{\Large \bf Gauge invariant formulation of massive totally symmetric

\bigskip
fermionic fields in  (A)dS space}

\vspace{2.5cm}

R.R. Metsaev\footnote{ E-mail: metsaev@lpi.ru }

\vspace{1cm}

{\it Department of Theoretical Physics, P.N. Lebedev Physical
Institute, \\ Leninsky prospect 53,  Moscow 119991, Russia }

\vspace{3.5cm}

{\bf Abstract}

\end{center}

\noindent Massive arbitrary spin totally symmetric free  fermionic
fields propagating in $d$-dimensional (Anti)-de Sitter space-time are
investigated. Gauge invariant action and the corresponding gauge
transformations for such fields are proposed.  The results are
formulated in terms of various mass parameters used in the literature
as well as the lowest eigenvalues of the energy operator. We  apply
our results to a study of partial masslessness of fermionic fields in
$(A)dS_d$,  and in  the case of $d=4$ confirm the conjecture made in
the earlier literature.

\newpage
\renewcommand{\thefootnote}{\arabic{footnote}}
\setcounter{footnote}{0}

\newsection{Introduction}

Conjectured duality \cite{Maldacena:1997re} of conformal ${\cal N}=4$
SYM theory and superstring theory in $AdS_5 \times S^5$ Ramond-Ramond
background has led to intensive study of field (string) dynamics in
AdS space. By now it is clear that in order to understand the
conjectured duality better it is necessary to develop powerful
approaches to study of field (string) dynamics in AdS space.
Light-cone approach is one of the promising approaches which might be
helpful to understand AdS/CFT duality better. As is well known,
quantization of Green-Schwarz superstrings propagating in flat space
is straightforward only in the light-cone gauge. Since, by analogy
with flat space, we expect that  quantization of the Green-Schwarz
$AdS$ superstring with  Ramond - Ramond flux \cite{Metsaev:1998it}
will be straightforward only in a light-cone gauge \cite{mt3} we
believe that from the stringy perspective of $AdS/CFT$ correspondence
the light-cone approach to field dynamics in $AdS$ is a fruitful
direction to go. Light-cone approach to dynamics of massive fields in
$AdS$ space was developed in \cite{Metsaev:1999ui,Metsaev:2003cu} and a
complete description of massive arbitrary spin bosonic and fermionic
fields in $AdS_5$ was obtained in \cite{Metsaev:2004ee}.

Unfortunately,  this is not enough for a complete study of the
$AdS/CFT$ correspondence because in order to apply the light-cone
approach to study of superstring in $AdS$ space we need  a light-cone
formulation of field dynamics in $AdS_5\times S^5$ Ramond-Ramond
background. Practically useful and self-contained way to give a
light-cone gauge description is to start with a Lorentz covariant and
gauge invariant description of field dynamics in $AdS_5\times S^5$
Ramond-Ramond background and then to impose the light-cone gauge. Our
experience led us to conclusion that the most simple way to develop
light-cone approach in $AdS_5\times S^5$  space is to start with
gauge invariant description of {\it fermionic fields}. It turns out,
however,  that gauge invariant description of {\it massive fermionic
fields} (with fixed but arbitrary spin) even in $AdS_5$ is still not
available in the literature. In this letter we develop Lagrangian
Lorentz covariant and gauge invariant formulation%
\footnote{Sometimes, a gauge invariant approach to massive fields is
referred to as a Stueckelberg approach.}
for {\it massive totally symmetric arbitrary spin fermionic fields}
in $(A)dS_d$ space. We believe that our results will be helpful to
find a  gauge invariant description of arbitrary spin fields in
$AdS_5\times S^5$ case. Our approach allows us to study fermionic
fields in $AdS_d$ space and $dS_d$ space on an equal footing. In this
letter we apply our results to study of partial masslessness of
fermionic fields in $(A)dS_d$. For $d=4$ our results confirm the
conjecture made in Ref.\cite{Deser:2001xr}.

Before proceeding  to the main theme of this letter let us mention
briefly the approaches which could be used to discuss gauge invariant
action for fields in $(A)dS$. Since the works
\cite{Fronsdal:1978vb}-\cite{Vasiliev:1987tk} devoted to massless
fields in $AdS_d$ various descriptions of massive and massless
arbitrary spin fields in $(A)dS$ have been developed. In particular,
an  ambient space formulation was discussed in
\cite{Metsaev:1995re}-\cite{Fotopoulos:2006ci} and various BRST
formulations were studied in
\cite{Buchbinder:2001bs}-\cite{Buchbinder:2006ge}. The frame-like
formulations of free fields which seems to be the most suitable for
formulation of the theory of interacting fields in $(A)dS$ was
developed in \cite{Alkalaev:2003qv,Alkalaev:2005kw}. Other
interesting formulations of higher spin theories were also discussed
recently in \cite{deMedeiros:2003px}-\cite{Baekler:2006vw}. In this
paper we adopt the approach of Ref.\cite{Zinoviev:2001dt} devoted to
the bosonic fields in $(A)dS$. This approach turns out to be the most
useful for our purposes.

\newsection{Gauge invariant action of massive fermionic field in
$(A)dS$}

In $d$-dimensional $(A)dS_d$ space the massive totally symmetric
arbitrary spin fermionic field is labelled by one mass parameter and
by one half-integer spin label $s+\frac{1}{2}$, where $s>0$ is an
integer number. To discuss Lorentz covariant and gauge invariant
formulation of such field we introduce Dirac complex-valued
tensor-spinor spin $s'+\frac{1}{2}$ fields of the $so(d-1,1)$ Lorentz
algebra $\psi^{A_1\ldots A_{s'}\alpha}$, $s'=0,1,\ldots, s$ (where
$A=0,1,\ldots, d-1$ are flat vector indices of the $so(d-1,1)$
algebra), i.e. we start with a collection of the tensor-spinor fields
\be \label{collect} \sum_{s'=0}^s \oplus\, \psi^{A_1\ldots
A_{s'}\alpha}\,. \ee
In order to obtain the gauge invariant description of a massive field
in an easy--to--use form, let us introduce a set of the creation and
annihilation operators $\alpha^A$, $\zeta$ and $\bar{\alpha}^A$,
$\bar{\zeta}$ defined by the relations%
\footnote{ We use oscillator formulation
\cite{Lopatin:1987hz,Labastida:1987kw} to handle the many indices
appearing for arbitrary spin fields. It can also be reformulated as
an algebra acting on the symmetric-spinor bundle on the manifold $M$
\cite{Hallowell:2005np}.
Note that the scalar oscillators $\zeta$, $\bar\zeta$ arise naturally
by a dimensional reduction \cite{Biswas:2002nk,Hallowell:2005np} from
flat space.
Our oscillators $\alpha^A$, $\bar\alpha^A$, $\zeta$, $\bar\zeta$ are
respective analogs of $dx^\mu$, $\partial_\mu$, $du$, $\partial_u$ of
Ref.\cite{Hallowell:2005np} dealing, among other thing, with massless
fermionic fields in $(A)dS$. We thank A.Waldron for pointing this to
us.}
\be\label{intver15} [\bar\alpha^A,\,\alpha^B]= \eta^{AB}\,,\qquad
[\bar{\zeta},\,\zeta] = 1\,,
\qquad \bar\alpha^A|0\rangle=0\,,\qquad \bar{\zeta}|0\rangle=0\,,\ee
where $\eta^{AB}$ is the mostly positive flat metric tensor. The
oscillators $\alpha^A$, $\bar\alpha^A$ and $\zeta$, $\bar\zeta$
transform in the respective vector and scalar representations of the
$so(d-1,1)$ Lorentz algebra. The tensor-spinor fields \rf{collect}
can be collected into a ket-vector $|\psi\rangle$ defined by
\beq
\label{intver16n1}
&& |\psi\rangle  \equiv \sum_{s'=0}^s
\zeta^{s-s'}|\psi_{s'}\rangle\,,
\\
&& \label{genfun2} |\psi_{s'}\rangle \equiv \alpha^{A_1}\ldots
\alpha^{A_{s'}} \psi^{A_1\ldots A_{s'}\alpha}(x)|0\rangle\,.
\eeq
Here and below spinor indices are implicit.  The ket-vector
$|\psi_{s'}\rangle$ \rf{genfun2} satisfies the constraint
\be
\label{homcon2} (\alpha^A\bar{\alpha}^A -s')|\psi_{s'} \rangle =0\,,
\qquad s'=0,1,\ldots, s\,,
\ee
which tells us that $|\psi_{s'}\rangle$ is a degree $s'$ homogeneous
polynomial in the oscillator $\alpha^A$. The tensor-spinor field
$|\psi_{s'}\rangle$ is subjected the basic algebraic constraint
\be \label{gamtra1}\gamma\bar\alpha \bar\alpha^2 |\psi_{s'} \rangle
=0\,, \quad \qquad s' = 0,1,\ldots, s\,, \qquad \gamma\bar\alpha
\equiv \gamma^A\bar\alpha^A\,,\qquad \bar\alpha^2 \equiv
\bar\alpha^A\bar\alpha^A\,,
\ee
which tells us that $|\psi_{s'} \rangle$ is a
reducible representation of the Lorentz algebra $so(d-1,1)$%
\footnote{Important constraint \rf{gamtra1} was introduced for the
first time in \cite{Fang:1979hq} while study of massless fermionic
fields in $AdS_4$. This constraint implies that the field
$|\psi_{s'}\rangle$ being reducible representation of the Lorentz
algebra $so(d-1,1)$ is decomposed into spin $s'+\frac{1}{2}$,
$s'-\frac{1}{2}$, $s'-\frac{3}{2}$ irreps of the Lorentz algebra.
Various Lagrangian formulations in terms of unconstrained fields in
flat space and (A)dS space may be found e.g. in
\cite{Buchbinder:2004gp}-\cite{Francia:2002pt}.}.
Note that for $s'=0,1,2$ the constraint \rf{gamtra1} is satisfied
automatically. In terms of the ket-vector $\psik$ \rf{intver16n1} the
algebraic constraints \rf{homcon2},\rf{gamtra1} take the form
\beq
&& \label{homcon2n} (N_\alpha  + N_\zeta  - s )|\psi\rangle =0\,,
\\
&& \label{gamtra1n} \gamma\bar\alpha \bar\alpha^2 |\psi\rangle =0\,,
\\
\label{Nzetadef} && N_\alpha \equiv \alpha^A\bar\alpha^A\,,\qquad
N_\zeta \equiv \zeta\bar\zeta\,. \eeq
Equation \rf{homcon2n} tells us that $|\psi\rangle$ is a degree $s$
homogeneous polynomial in the oscillators $\alpha^A$, $\zeta$.

Lagrangian for the massive fermionic field in $(A)dS_d$ space we
found takes the form
\be\label{sec5005} \LL = \LL_{der} + \LL_m\,, \ee
where $\LL_{der}$ stands for a derivative depending part of $\LL$,
while $ \LL_m$ stands for a mass part of $\LL$:%
\footnote{$\psibr$ is defined according the rule $\psibr =
(\psik)^\dagger\gamma^0$. We use $e\equiv \det e_\mu^A$, where
$e_\mu^A$ is vielbein of $(A)dS_d$ space.}
\beq
&& \label{sec5006} {\rm i}e^{-1} \LL_{der} =   \psibr L \psik\,,
\\[5pt]
&&  \label{sec5007}   {\rm i}e^{-1} \LL_m  =  \psibr \MM \psik\,.
\eeq
The standard first-derivative differential operator $L$ which enters
$\LL_{der}$ \rf{sec5006} is given by
\beq
\label{Ldef} && L \equiv \Dline  - \alpha D \gamma\bar\alpha -
\gamma\alpha\,\bar\alpha D + \gamma\alpha \Dline\gamma\bar\alpha +
\frac{1}{2}\gamma\alpha\,\alpha D \bar\alpha^2 +
\frac{1}{2}\alpha^2\gamma\bar\alpha\,\bar\alpha D -
\frac{1}{4}\alpha^2\Dline\bar\alpha^2\,, \ \ \ \ \eeq
where we use the notation
\be \gamma\alpha \equiv  \gamma^A\alpha^A\,, \quad \gamma\bar\alpha
\equiv \gamma^A\bar\alpha^A\,, \quad \alpha^2 \equiv \alpha^A
\alpha^A\,,\quad \bar\alpha^2\equiv\bar\alpha^A \bar\alpha^A \,,\ee
\be \label{vardef01}
\Dline \equiv \gamma^A D^A\,,\qquad \alpha D \equiv \alpha^A
D^A\,,\qquad \bar\alpha D \equiv \bar\alpha^A D^A\,, \qquad
D_A \equiv e_A^\mu D_\mu\,, \ee
and $e_A^\mu$ stands for inverse vielbein of $(A)dS_d$ space, while
$D_\mu$ stands for the Lorentz covariant derivative
\be \label{lorspiope} D_\mu \equiv
\partial_\mu
+\frac{1}{2}\omega_\mu^{AB}M^{AB}\,.\ee
The $\omega_\mu^{AB}$ is the Lorentz connection of $(A)dS_d$ space,
while a spin operator $M^{AB}$ forms a representation of the Lorentz
algebra $so(d-1,1)$:
\be\label{loralgspiope} M^{AB} = M_b^{AB} +
\frac{1}{2}\gamma^{AB}\,,\qquad M_b^{AB} \equiv \alpha^A \bar\alpha^B
- \alpha^B \bar\alpha^A\,,  \qquad \gamma^{AB} \equiv
\frac{1}{2}(\gamma^A\gamma^B - \gamma^B\gamma^A)\,.\ee
We note that our derivative depending part of the Lagrangian
$\LL_{der}$ \rf{sec5006} is nothing but a sum of the Lagrangians of
Ref.\cite{Fang:1979hq} for the tensor-spinor fields \rf{collect}.

We now proceed with discussion of the mass operator $\MM$
\rf{sec5007}. The operator $\MM$ is given by
\beq
\label{sec5008} \MM &= & (1 - \gamma\alpha \gamma\bar\alpha
-\frac{1}{4} \alpha^2 \bar\alpha^2 )\mwt_1 + \mwt_4 (\gamma\alpha
\bar\zeta - \frac{1}{2} \alpha^2 \bar\zeta \gamma\bar\alpha)
\nonumber\\
& - & (\zeta  \gamma\bar\alpha  - \frac{1}{2}\gamma\alpha \zeta
\bar\alpha^2)\mwt_4\,, \eeq
where operators $\mwt_1$, $\mwt_4$ do not depend on the
$\gamma$-matrices and $\alpha$-oscillators, and take the form
\beq
\label{mwt1sol} && \mwt_1  = \frac{2s+ d -2}{2s + d - 2 - 2N_\zeta}
\mas \,,
\\[3pt]
\label{mwt4sol} && \mwt_4  = \Bigl(\frac{2s+ d -3 - N_\zeta}{2s + d -
4 - 2N_\zeta}\, F(\mas,s,N_\zeta)\Bigr)^{1/2} \,.
\eeq
Function $F(\mas,s,N_\zeta)$ depends on a mass parameter $\mas$, spin
$s$ and operator $N_\zeta$, and is given by
\beq
\label{Fdef} && F(\mas,s,N_\zeta) = \mas^2 + \coscon \Bigl(s +
\frac{d-4}{2} - N_\zeta\Bigr)^2  \,.
\eeq
$F$ is restricted to be positive and throughout this paper, unless
otherwise specified, we use the convention:%
\footnote{Thus our Lagrangian gives description of massive fermionic
fields in $(A)dS$ space and flat space on an equal footing.
Discussion of massive fermionic fields in flat space in framework of
BRST approach may be found in
\cite{Buchbinder:2004gp,Buchbinder:2006nu}}
\be \label{omegadef}
\coscon = \left\{\begin{array}{cl}
-1 & \hbox{for AdS space},
\\
0 & \hbox{for flat space},
\\
+1 & \hbox{for dS space}.
\end{array}\right.
\ee
The mass parameter $\mas$ is a freedom of our solution, i.e. gauge
invariance allows us to find Lagrangian completely by module of mass
parameter as it should be for the case of massive fields.

Now we discuss gauge symmetries of the action
\be\label{action} S = \int d^d x\, \LL\,. \ee
To this end we introduce parameters of gauge transformations
$\epsilon^{A_1\ldots A_{s'}\alpha}$, $s'=0,1,\ldots, s-1$ which are
$\gamma$-traceless (for $s'>0$) Dirac complex-valued tensor-spinor
spin $s'+\frac{1}{2}$ fields of the $so(d-1,1)$ Lorentz algebra, i.e.
we start with a collection of the tensor-spinor fields
\be \label{epscollect} \sum_{s'=0}^{s-1} \oplus\, \epsilon^{A_1\ldots
A_{s'}\alpha}\,,\qquad \quad \gamma^A \epsilon^{AA_2\ldots
A_{s'}}=0\,, \quad \hbox{for }\ \  s'>0\,. \ee
As before to simplify our expressions we use the ket-vector of gauge
transformations parameter
\beq
\label{gaugpar1}
&& \epsilonk  \equiv \sum_{s'=0}^{s-1}
\zeta^{s-1-s'}|\epsilon_{s'}\rangle\,,
\\
&& \label{gaugpar2} |\epsilon_{s'}\rangle \equiv \alpha^{A_1}\ldots
\alpha^{A_{s'}} \epsilon^{A_1\ldots A_{s'}\alpha}(x)|0\rangle\,.
\eeq
The ket-vector $\epsilonk$ satisfies the algebraic constraints
\beq
&& \label{gaugpar3} (N_\alpha  + N_\zeta  - s + 1 )\epsilonk =0\,,
\\
&& \label{gaugpar4} \gamma\bar\alpha \epsilonk =0\,.
\eeq
The constraint \rf{gaugpar3} tells us that the ket-vector $\epsilonk$
is a degree $s-1$ homogeneous polynomial in the oscillators
$\alpha^A$, $\zeta$, while the constraint \rf{gaugpar4} respects the
$\gamma$-tracelessness of $\epsilonk$.

Now the gauge transformations under which the action \rf{action} is
invariant take the form
\beq
&& \label{gaugtrapsi} \delta \psik = (\alpha D + \FF)\epsilonk \,,
\\[3pt]
&& \FF \equiv \zeta \fwt_1 + \gamma\alpha\fwt_2 + \alpha^2 \fwt_3
\bar\zeta \,, \eeq
where operators $\fwt_1$, $\fwt_2$, $\fwt_3$ do not depend on the
$\gamma$-matrices and $\alpha$-oscillators, and take the form
\beq
\label{Del1def} && \fwt_1  = \Bigl(\frac{2s+ d -3 - N_\zeta}{2s + d -
4 - 2N_\zeta}\, F(\mas,s,N_\zeta)\Bigr)^{1/2}\,,
\\[9pt]
\label{Del2def}  && \fwt_2  =   \frac{2s+ d -2}{(2s + d - 2 -
2N_\zeta)(2s + d - 4 - 2N_\zeta)} \mas \,,
\\[9pt]
\label{Del3def}  && \fwt_3  = -\Bigl(\frac{2s+ d -3 - N_\zeta}{ (2s +
d - 4 - 2N_\zeta)^3}\, F(\mas,s,N_\zeta)\Bigr)^{1/2}\,,
\eeq
and $F$ is defined in \rf{Fdef}. Thus we expressed our results in
terms of the mass parameter $\mas$. Since there is no commonly
accepted definition of mass in $(A)dS$ we relate our mass parameter
$\mas$ with various mass parameters used in the literature. One of
the most-used definitions of mass, which we denote by $m_\smD$, is
obtained from the following expansion of mass part of the Lagrangian:
\be\label{mDdef} {\rm i}e^{-1} \LL_m = \langle \psi_s | m_\smD
|\psi_s \rangle + \ldots \,,\ee
where dots stand for terms involving $|\psi_{s'}\rangle$, $s'<s$, and
for contribution which vanishes while imposing the constraint
$\gamma\bar\alpha|\psi\rangle$. Comparing \rf{mDdef} with
\rf{sec5008},\rf{mwt1sol} leads then to the identification
\be\label{kappamD} \mas = m_\smD\,.\ee
Another definition of mass parameter for fermionic fields in $AdS_d$
\cite{Metsaev:2003cu}, denoted by $m$, can be obtained by requiring
that the value of $m=0$ corresponds to the massless fields. For the
case of spin $s+\frac{1}{2}$ field in $AdS_d$ the mass parameter $m$
is related with $m_\smD$ as
\be \label{mDmrel} m_\smD  = m + s + \frac{d-4}{2}\qquad \qquad
\hbox{ for } \ AdS_d\,,\ee
where $m>0$ corresponds to massive unitary irreps of the $so(d-1,2)$
algebra \cite{Metsaev:1995re,Metsaev:2003cu}. Below we demonstrate
that natural generalization of \rf{mDmrel} which is valid for both
$AdS$ and $dS$ spaces is given by
\be \label{mDmrelnew} m_\smD  = m + \sqrt{- \coscon }(s +
\frac{d-4}{2}) \qquad\quad \hbox{ for } \ (A)dS_d \,.\ee
Since sometimes in the case of $AdS$ the formulation in terms of the
lowest eigenvalue of energy operator $E_0$ is preferable we now
express our results in terms of $E_0$. To this end we use the
relation found in \cite{Metsaev:2003cu}:
\be m = E_0 - s - d +\frac{5}{2} \qquad \qquad \hbox{ for } \ AdS_d
\,.\ee
Making use then \rf{kappamD},\rf{mDmrel} we get for the case of
$AdS_d$ the desired relations
\beq
&& \mas = E_0 - \frac{d-1}{2}\,,
\\
&& F =  (E_0 - s -d + \frac{5}{2} + N_\zeta)(E_0 + s -\frac{3}{2} -
N_\zeta)  \,.
\eeq

\newsection{Limit of massless fields in $(A)dS_d$}

In previous section we presented the action for the massive field. In
limit as the mass parameter $m$ tends to zero our Lagrangian leads to
the Lagrangian for massless field  in $(A)dS_d$. Let us discuss the
massless limit in detail. To realize limit of massless field in
$(A)dS_d$ we take (see \rf{kappamD},\rf{mDmrelnew})
\be\label{m0lim} \mas  \rightarrow  \sqrt{- \coscon }(s +
\frac{d-4}{2}) \quad \Longleftrightarrow \quad m \rightarrow 0\,. \ee
We now demonstrate that this limit leads to appearance of the
invariant subspace in $|\psi\rangle$ \rf{intver16n1} and this
invariant subspace, denoted by $|\psi^{m=0}\rangle$, is given by the
leading ($s'=s$) term in \rf{intver16n1}:
\be \label{psimzero} |\psi^{m=0}\rangle = |\psi_s\rangle\,. \ee
All that is required is to demonstrate that in the limit \rf{m0lim},
the ket-vector $|\psi^{m=0}\rangle$ satisfies the following
requirements: i) $|\psi^{m=0}\rangle$ is invariant under action of
the mass operator $\MM$; ii) the gauge transformation of the
ket-vector $|\psi^{m=0}\rangle$ becomes the standard gauge
transformation of massless field. To this end we note that an action
of the mass operator $\MM$ on $|\psi_s\rangle$ and the gauge
transformation of $|\psi_s\rangle$ take the form
\beq
&& \label{MMaction} \MM |\psi_s\rangle =  (1- \gamma\alpha
\gamma\bar\alpha - \frac{1}{4} \alpha^2
\bar\alpha^2)\mwt_1(0)|\psi_s\rangle + (\gamma\alpha -\frac{1}{2}
\alpha^2  \gamma\bar\alpha)\mwt_4(0)|\psi_{s-1}\rangle\,,
\\[5pt]
&& \label{gaugtranpsis} \delta |\psi_s\rangle = (\alpha D  +
\gamma\alpha \fwt_2(0))|\epsilon_{s-1}\rangle + \alpha^2 \fwt_3(0)
|\epsilon_{s-2}\rangle\,,
\eeq
where $\mwt_{1,4}(0)$ \rf{MMaction} and $\fwt_{2,3}(0)$
\rf{gaugtranpsis} stand for $\mwt_{1,4}$ \rf{mwt1sol},\rf{mwt4sol}
and $\fwt_{2,3}$ \rf{Del2def}, \rf{Del3def} in which we set
$N_\zeta=0$.
Taking into account
\be
\label{limMMaction} \lim_{m\rightarrow 0}\,  \mwt_4(0) = 0\,, \qquad
\lim_{m\rightarrow 0}\, \fwt_3(0) = 0\,,
\ee
we see that if $m = 0$ then the ket-vector $|\psi_s\rangle$ is indeed
invariant under action of the mass operator $\MM$ and a realization
of the mass operator on the ket-vector $|\psi_s\rangle$ takes the
form

\be \label{MMmassdef} \MM^{m=0} = \sqrt{- \coscon } (s +
\frac{d-4}{2})(1 - \gamma\alpha \gamma\bar\alpha - \frac{1}{4}
\alpha^2 \bar\alpha^2)\,,\ee
while the relations \rf{gaugtranpsis}, \rf{limMMaction} lead to the
gauge transformation
\be
\delta |\psi^{m=0}\rangle = (\alpha D  + \frac{\sqrt{- \coscon }}{2}
\gamma\alpha)|\epsilon_{s-1}\rangle\,,
\ee
which is noting but the standard gauge transformation of massless
field in $(A)dS$ space. Thus the Lagrangian for massless spin
$s+\frac{1}{2}$ fermionic
field in $(A)dS_d$ space takes the form%
\footnote{Our Lagrangian \rf{ourLag} is a generalization to
$d$-dimensions of the Lagrangian of Ref.\cite{Fang:1979hq} for
massless field in $(A)dS_4$. Alternative Lagrangian descriptions of
massless fermionic fields in $(A)dS_d$ may be found in
\cite{Vasiliev:1987tk,Hallowell:2005np}}
\be\label{ourLag} {\rm i}e^{-1} \LL =   \langle\psi^{m=0}| L +
\MM^{m=0} |\psi^{m=0}\rangle \,, \ee
where $|\psi^{m=0}\rangle$ is given by \rf{psimzero},\rf{genfun2},
while the operators $L$ and $\MM^{m=0}$ are defined by \rf{Ldef} and
\rf{MMmassdef} respectively. The remaining ket-vectors
$|\psi_{s-1}\rangle,\ldots, |\psi_0\rangle$ in \rf{intver16n1}
decouple in the massless limit and they describe spin $s-\frac{1}{2}$
massive field, i.e. in the massless limit the generic field $\psik$
is decomposed into two decoupling systems - one massless spin
$s+\frac{1}{2}$ field and one massive spin $s-\frac{1}{2}$ field.
Adopting \rf{mDdef} for spin $s-\frac{1}{2}$ field we find mass of
the massive spin $s-\frac{1}{2}$ field: $\mas=\sqrt{- \coscon
}(s+(d-2)/2)$.

\newsection{ Partial masslessness of fermionic fields in $(A)dS_d$}

Here we apply our results to study of partial masslessness%
\footnote{ Partial masslessness was discovered in
\cite{Deser:1983mm}. Recent discussion of this theme and to some
extent complete  list of references may be found in
\cite{Skvortsov:2006at}.}
of fermionic fields in $(A)dS_d$. We confirm conjecture of
Ref.\cite{Deser:2001xr} for $d=4$ and obtain an generalization to the
case of arbitrary $d>4$. In this section we assume that the $\coscon$
\rf{omegadef} takes the values $\pm1$. We start our discussion of
partial masslessness of fermionic fields with simplest case of (see
also Ref.\cite{Deser:2001us})

{\bf Massive spin $5/2$ field}. Such field is described by
ket-vectors $|\psi_2\rangle$, $|\psi_1\rangle$, $|\psi_0\rangle$ (see
\rf{intver16n1}). For spin $5/2$  field there is one critical value
of $\mas$ which leads to appearance of partial massless field. For
this critical value of $\mas$ the generic field $\psik$ is decomposed
into one partial massless field and one massive spin $\frac{1}{2}$
field. To demonstrate this we consider the gauge transformations
\rf{gaugtrapsi},
\beq
&& \delta  |\psi_2\rangle = \alpha D |\epsilon_1\rangle +
\gamma\alpha \fwt_2(0) |\epsilon_1\rangle + \alpha^2
\fwt_3(0)|\epsilon_0\rangle\,,
\\[5pt]
&& \delta  |\psi_1\rangle = \alpha D |\epsilon_0\rangle + \fwt_1(0)
|\epsilon_1\rangle + \gamma\alpha \fwt_2(1)
 |\epsilon_0\rangle\,,
\\[5pt]
&& \delta  |\psi_0\rangle =  \fwt_1(1)|\epsilon_0\rangle\,,
\eeq
where $\fwt_{1,2,3}(n)$ are given in \rf{Del1def}-\rf{Del3def} in
which we set $s=2$ and argument $n$ stands for an eigenvalue of the
operator $N_\zeta$. The critical value of $\mas$ is obtained from the
requirement of decoupling of the field $|\psi_0\rangle$. This
requirement amounts to the equation $\fwt_1(1)=0$ which leads to the
critical value
\be \mas_{(0)}^2 = - \coscon  \Bigl(1 + \frac{d-4}{2}\Bigr)^2\,.\ee
For this value of $\mas$ the generic field $|\psi\rangle$ is
decomposed into two decoupling systems -- one partial massless field
described by $|\psi_2\rangle$, $|\psi_1\rangle$ and one massive spin
$\frac{1}{2}$ field described by $|\psi_0\rangle$. We proceed with
discussion of partial masslessness for

{\bf Massive spin $7/2$ field}. Spin $7/2$ field $\psik$ is described
by ket-vectors $|\psi_3\rangle$, $|\psi_2\rangle$, $|\psi_1\rangle$,
$|\psi_0\rangle$ (see \rf{intver16n1}). The gauge transformations
\rf{gaugtrapsi} for these ket-vectors take the form
\beq
&& \delta  |\psi_3\rangle = \alpha D |\epsilon_2\rangle +
\gamma\alpha \fwt_2(0) |\epsilon_2\rangle + \alpha^2
\fwt_3(0)|\epsilon_1\rangle\,,
\\[5pt]
\label{7/2gautr2} && \delta  |\psi_2\rangle = \alpha D
|\epsilon_1\rangle + \fwt_1(0)|\epsilon_2\rangle + \gamma\alpha
\fwt_2(1) |\epsilon_1\rangle + 2\alpha^2
\fwt_3(1)|\epsilon_0\rangle\,,
\\[5pt]
\label{7/2gautr1}  && \delta  |\psi_1\rangle = \alpha D
|\epsilon_0\rangle + \fwt_1(1) |\epsilon_1\rangle + \gamma\alpha
\fwt_2(2)
 |\epsilon_0\rangle\,,
\\[5pt]
\label{7/2gautr0} && \delta  |\psi_0\rangle =
\fwt_1(2)|\epsilon_0\rangle\,,
\eeq
where expressions for $\fwt_{1,2,3}(n)$ are given in
\rf{Del1def}-\rf{Del3def} in which we set $s=3$ and argument $n$
stands for an eigenvalue of the operator $N_\zeta$. For the spin
$7/2$ field there are two critical values of $\mas$. For each
critical value of $\mas$ the generic field $\psik$ is decomposed into
one partial massless field and one massive field. We consider these
critical values in turn.

First critical value of $\mas$ is obtained from the requirement of
decoupling of the field $|\psi_0\rangle$ (see \rf{7/2gautr0}). This
requirement amounts to the equation $\fwt_1(2)=0$ which leads to the
critical value
\be \mas_{(0)}^2 = - \coscon  \Bigl(1 + \frac{d-4}{2}\Bigr)^2\,.\ee
For this value of $\mas$ the generic field $|\psi\rangle$ is
decomposed into two decoupling systems - one partial massless field
described by $|\psi_3\rangle$, $|\psi_2\rangle$, $|\psi_1\rangle$ and
one massive spin $\frac{1}{2}$ field $|\psi_0\rangle$.

The second critical value of $\mas$ is obtained from the requirement
of decoupling of the fields $|\psi_1\rangle$, $|\psi_0\rangle$ (see
\rf{7/2gautr2},\rf{7/2gautr1}). This requirement amounts to the
equations $\fwt_1(1)=0$, $\fwt_3(1)=0$ which lead to the critical
value
\be \mas_{(1)}^2 = - \coscon  \Bigl(2 + \frac{d-4}{2}\Bigr)^2\,.\ee
For this $\mas$ the generic field $|\psi\rangle$ is decomposed into
one partial massless field described by $|\psi_3\rangle$,
$|\psi_2\rangle$ and one massive spin $\frac{3}{2}$ field described
by $|\psi_1\rangle$, $|\psi_0\rangle$. We finish with partial
masslessness for

{\bf Massive arbitrary spin $s+\frac{1}{2}$ field}. Such field is
described by ket-vectors $|\psi_{s'}\rangle$, $s'=0,1,\ldots s$.
Gauge transformations \rf{gaugtrapsi} for these ket-vectors take the
form
\beq
\label{psistgautra} \delta  |\psi_{s'}\rangle & = & \alpha D
|\epsilon_{s'-1}\rangle + \fwt_1(s-s'-1) |\epsilon_{s'}\rangle
\nonumber\\[5pt]
& + & \gamma\alpha\fwt_2(s-s') |\epsilon_{s'-1}\rangle +
(s-s'+1)\alpha^2 \fwt_3(s-s')|\epsilon_{s'-2}\rangle\,,
\eeq
where $\fwt_{1,2,3}(n)$ are given in \rf{Del1def}-\rf{Del3def} and
argument $n$ stands for an eigenvalue of the operator $N_\zeta$. For
values $s'=0,1,s$ \rf{psistgautra} we use the convention
$|\epsilon_{-2}\rangle = |\epsilon_{-1}\rangle = |\epsilon_{s}
\rangle = 0$. For the spin $s+\frac{1}{2}$ field there are $s-1$
critical values of $\mas$, denoted by $\mas_{(n)}$, $n=0,1,\ldots,
s-2$ (the case of $n=s-1$ leads to massless field and was considered
in Section 3). For each $\mas_{(n)}$ the generic field $\psik$ is
decomposed into one partial massless field and one spin
$n+\frac{1}{2}$ massive field. To find $\mas_{(n)}$ we note that the
requirement of decoupling of the fields $|\psi_n\rangle,\ldots,
|\psi_0\rangle$, $n=0,\ldots,s-2$ amounts to equations
$\fwt_1(s-n-1)=0$, $\fwt_3(s-n-1)=0$. Solution to these
equations%
\footnote{ We note that the key point is not positivity or even
reality of the mass part of action \rf{sec5007}, but rather stability
of the energy and unitarity of the underlying physical
representations. For bosons a negative mass term is allowed in $AdS$
(the Breitenlohner-Freedman bound, \cite{Breitenlohner:1982bm}),
while partially massless fermions even have an imaginary mass term in
their actions but are still stable and unitary in $dS$. Partially
massless fermions are not unitary in $AdS$ (see
Refs.\cite{Deser:2001xr,Deser:2001us,Deser:2003gw}).}
\be \label{mdnsol} \mas_{(n)}^2 = - \coscon  \Bigl(n+1 +
\frac{d-4}{2}\Bigr)^2\,,\ee
is in agreement with conjecture made in Ref.\cite{Deser:2001xr} for
the case of $d=4$. Thus we confirmed conjecture of
Ref.\cite{Deser:2001xr} and obtained $\mas_{(n)}$ for $d>4$. For each
$\mas_{(n)}$ the generic field $|\psi\rangle$ is decomposed into two
decoupling systems - one partial massless field
$|\psi_{par}^{(s)}\rangle$ described by $|\psi_s\rangle,\ldots,
|\psi_{n+1}\rangle$, and one massive spin $n+\frac{1}{2}$ field
$|\psi_{msv}^{(n)}\rangle$ described by $|\psi_n\rangle,\ldots,
|\psi_0\rangle$. This is to say that by decomposing $\psik$
\rf{intver16n1} into the respective ket-vectors
\be\label{vardef22}
|\psi_{par}^{(s)}\rangle  \equiv \sum_{s'=n+1}^s
\zeta^{s-s'}|\psi_{s'}\rangle\,,
\qquad
|\psi_{msv}^{(n)}\rangle  \equiv \sum_{s'=0}^n
\zeta^{s-s'}|\psi_{s'}\rangle\,,
\ee
one can make sure that if $\mas = \mas_{(n)}$ then the mass part of
the Lagrangian \rf{sec5007} is factorized
\be {\rm i}e^{-1} \LL_m = \langle \psi_{par}^{(s)}|
\MM|\psi_{par}^{(s)}\rangle + \langle \psi_{msv}^{(n)}|
\MM|\psi_{msv}^{(n)}\rangle\,. \ee
For values $\mas = \mas_{(n)}$ the gauge transformations
\rf{gaugtrapsi} are also factorized, while $\LL_{der}$ \rf{sec5006}
is factorized for arbitrary $\mas$%
\footnote{ $\mas$- and $m$- masses of the field
$|\psi_{par}^{(s)}\rangle$ are given by: $\mas = \sqrt{- \coscon
}(n+1 +(d-4)/2)$, $m= \sqrt{- \coscon }(n+1-s)$, while for the field
$|\psi_{msv}^{(n)}\rangle$ we get $\mas=\sqrt{- \coscon
}(s+(d-2)/2)$, $m=\sqrt{- \coscon }(s -n + 1)$.}.

\newsection{ Uniqueness of Lagrangian for massive fermionic field}

We now demonstrate that the Lagrangian and gauge transformations are
uniquely determined by requiring that the action be gauge invariant.
We formulate our statement. Suppose the derivative depending part of
the Lagrangian is given by \rf{sec5006}, while the derivative
depending part of gauge transformations \rf{gaugtrapsi} is governed
by $\alpha D$-term. Suppose the gauge field $\psik$ \rf{intver16n1}
and the gauge transformations parameter $\epsilonk$ \rf{gaugpar1}
satisfy the respective constraints \rf{gamtra1n},\rf{gaugpar4}. Then
we state that the mass operator $\MM$ given in \rf{sec5008} and the
operator $\FF$ which enters gauge transformations \rf{gaugtrapsi} are
uniquely determined by the following requirements: i) the action be
gauge invariant; ii) there are no invariant subspaces in $\psik$
under action of gauge transformations. Here we outline prove of this
statement. We start with general form of the mass operator $\MM$ and
the operator $\FF$:
\beq
\label{sec5008nn} \MM &= & m_1 +  \gamma\alpha m_2 \gamma\bar\alpha +
\alpha^2 m_3 \bar\alpha^2  + \gamma\alpha m_4 + \alpha^2 m_5
\gamma\bar\alpha
\nonumber\\[5pt]
&-& m_4^\dagger \gamma\bar\alpha  - \gamma\alpha m_5^\dagger
\bar\alpha^2
 + \alpha^2m_6 + m_6^\dagger \bar\alpha^2 \,,
\\[5pt]
\label{sec5008nn11}  \FF & = &  f_1 + \gamma\alpha f_2 + \alpha^2
f_3\,,
\eeq
where $m_{1,\ldots,6}$ and $f_{1,2,3}$ do depend on $\gamma$-matrices
and $\alpha$-oscillators, and are given by
\beq &
m_1 = \mwt_1 \,, \quad
m_2 =  \mwt_2 \,, \quad
m_3 =  \mwt_3 \,, \quad
m_4 =  \mwt_4 \bar\zeta\,, \quad
m_4^\dagger  =  \zeta \mwt_4^\dagger \,,
&
\\[5pt]
& m_5 =  \mwt_5 \bar\zeta\,, \quad
m_5^\dagger =  \zeta \mwt_5^\dagger\,, \quad
m_6 =  \mwt_6 \bar\zeta^2\,, \quad
m_6^\dagger =  \zeta^2 \mwt_6^\dagger\,,
&
\\[5pt]
\label{DelDelwt} & f_1 = \zeta \fwt_1 \,,\qquad
f_2 =  \fwt_2 \,, \qquad
f_3 =  \fwt_3 \bar\zeta \,. & \eeq
Operators $\mwt_{1,\ldots,6}$ and $\fwt_{1,2,3}$ depend only on
$N_\zeta$ \rf{Nzetadef}. $\mwt_{1,\ldots,6}^\dagger$ stand for
hermitian conjugated of $\mwt_{1,\ldots,6}$. Since $\mwt_{1,2,3}$ are
hermitian, $\mwt_{1,2,3}^\dagger = \mwt_{1,2,3}$, and depend only on
the hermitian operator $N_\zeta$ the operators $\mwt_{1,2,3}$ are
real-valued functions of $N_\zeta$ from the very beginning. These
properties of $\mwt_{1,\ldots,6}$ and $\fwt_{1,2,3}$ and expressions
for $\MM$ \rf{sec5008nn}, $\FF$ \rf{sec5008nn11} are obtained by
requiring that:

\noindent {\bf i}) $\MM$ and $\FF$ commute with the spin operator of
the Lorentz algebra $M^{AB}$ \rf{loralgspiope} and satisfy the
commutators $[N_\alpha + N_\zeta,\MM]=0$, $[N_\alpha +
N_\zeta,\FF]=\FF$;

\noindent {\bf ii}) $\MM$ does not involve terms like
$\alpha^2\gamma\alpha p_1 $ and $p_2\bar\alpha^2\gamma\bar\alpha$,
where $p_{1,2}$ are polynomial in the oscillators (such terms in view
of \rf{gamtra1n} do not contribute to $\LL_m$);

\noindent {\bf iii}) $\FF$ does not involve terms like
$\alpha^2\gamma\alpha p_3$, $p_4 \gamma\bar\alpha$, where $p_{3,4}$
are polynomial in the oscillators (the $p_3$-terms lead to violation
of constraint \rf{gamtra1n} for gauge transformed field
\rf{gaugtrapsi}, while the $p_4$-terms in view of \rf{gaugpar4} do
not contribute to $\delta\psik$ \rf{gaugtrapsi});

\noindent {\bf iv}) $\MM$ and hermitian conjugated of $\MM$ satisfy
the relation $\MM^\dagger =-\gamma^0 \MM \gamma^0$.

Thus all that is required is to find dependence of the operators
$\mwt_{1,\ldots,6}$ and $\fwt_{1,2,3}$ on $N_\zeta$. We now
demonstrate that this dependence can be determined by requiring that
the action be gauge invariant. We evaluate the variation of the
action \rf{sec5005},\rf{action} under gauge transformations
\rf{gaugtrapsi},\rf{sec5008nn11},
\beq
\delta S &=& -{\rm i} \int d^dx\, e \psibr\Bigl(
\Dline X_1
+ \gamma\alpha\,\bar\alpha D X_2
+ \alpha D X_3
+ \gamma\alpha \Dline X_4
+ \alpha^2\bar\alpha D X_5
\nonumber\\[5pt]
&& + \alpha^2 \Dline X_6
+ \alpha D \gamma\alpha X_7 + \alpha^2\alpha D X_8 + \bar\alpha D X_9
\nonumber\\[5pt]
&&+  Y^\smzero  + \gamma\alpha (Y^\smone +
Y_{\scriptscriptstyle(A)dS}^\smone )+ \alpha^2 Y^\smtwo
\Bigr)\epsilonk + h.c\,, \eeq
where we use the notation
\beq
&& X_1 = f_1 -  m_4^\dagger\,, \qquad
X_2 = -f_1 - 2 m_5^\dagger\,,
\\[5pt]
&& X_3 = -(2s + d - 4 -2N_\zeta)f_2 + m_1\,,
\\[5pt]
&& X_4 = (2s + d - 4 -2N_\zeta) f_2 +  m_2\,,
\\[5pt]
&& X_5 = \frac{1}{2}(2s + d - 4 -2N_\zeta) f_2 + 2m_3\,,
\\[5pt]
&& X_6 = -\frac{1}{2}(2s + d - 4 -2N_\zeta) f_3 + m_5 \,,
\\[5pt]
&& X_7 = (2s + d - 4 -2N_\zeta) f_3 + m_4\,,
\\[5pt]
&& X_8 = m_6\,, \qquad \
X_9 = 2m_6^\dagger\,,
\eeq
\beq \label{Yzerodef}
Y^\smzero & = &m_1 f_1 - (2s + d - 2 N_\zeta)m_4^\dagger f_2\,,
\\[5pt]
Y^\smone & = & m_1 f_2 + (2s + d - 2 -2N_\zeta)m_2 f_2 + m_4 f_1
\nonumber\\[4pt]
&  - & 2m_4^\dagger f_3 - 2(2s + d - 2 -2N_\zeta)m_5^\dagger f_3\,, \
\ \ \
\\[5pt]
Y^\smtwo & = & m_1 f_3 + 2m_2 f_3 + 2(2s+ d - 4 - 2 N_\zeta ) m_3 f_3
\nonumber\\[5pt]
&+& m_4 f_2 + (2s + d - 4 -2N_\zeta)m_5 f_2\,,
\\
[5pt]
\label{YAdSdef}
Y_{\scriptscriptstyle(A)dS}^\smone & = & -\frac{\coscon}{4}( 2s + d -
3 - 2 N_\zeta)( 2s + d - 4 - 2 N_\zeta)\,.
\eeq
Requiring this variation to vanish gives the equations%
\footnote{ Expressions for $Y$'s given in \rf{Yzerodef}-\rf{YAdSdef}
are valid provided that the equations $X_8=X_9=0$ are satisfied.}
\beq
&& \label{Xaequ} X_a= 0\,,\qquad a =1,\ldots, 9\,,
\\[3pt]
\label{Y0equ} && Y^\smzero =0\,,
\\[3pt]
\label{Y1equ}&& Y^\smone + Y_{\scriptscriptstyle (A)dS}^\smone = 0
\,,
\\[3pt]
\label{Y2equ}&& Y^\smtwo = 0\,. \eeq
Solution to Eqs.\rf{Xaequ} is easily found to be
\beq
\label{msol1} && m_1 = (2s + d - 4 -2N_\zeta) f_2\,, \qquad
 m_4 = -(2s + d - 4 -2N_\zeta) f_3\,,
\\[5pt]
\label{msol2} && m_2 = - m_1\,,\quad m_3 = -\frac{1}{4} m_1\,,\quad
m_5 = -\frac{1}{2}m_4\,,
\\[5pt]
\label{msol3}&& m_4^\dagger = f_1\,,
\qquad
m_5^\dagger = -\frac{1}{2} f_1\,,
\qquad
m_6 = m_6^\dagger =0\,, \eeq
i.e. Eqs.\rf{Xaequ} allow us to express the operators $m_a$,
$a=1,\ldots,5$, entirely in terms of the operators $f_2$, $f_3$ which
enter the gauge transformations. Moreover, the expressions for $m_4$
\rf{msol1} and $m_4^\dagger$ \rf{msol3} imply the relation
\be \label{deldel3} f_1^\dagger = -(2s + d - 4 -2N_\zeta) f_3\,.\ee
We now analyze \rf{Y0equ}. Inserting $m_1$ \rf{msol1} and
$m_4^\dagger$ \rf{msol3} in \rf{Y0equ} we cast \rf{Y0equ} in the form
\be \label{sec5024n2} f_1 \Bigl( (2s+d-6-2N_\zeta) f_2^+
-(2s+d-2-2N_\zeta) f_2\Bigr)  =0\,,
\ee
where $f_2 \equiv f_2(N_\zeta)$ and $f_2^+ \equiv f_2(N_\zeta+1)$.
Solution to \rf{sec5024n2} $f_1=0$ for all $N_\zeta$ (or $f_1=0$ for
some particular eigenvalues of $N_\zeta$) leads to massless fields
(or partial massless fields), i.e. such solution leads to appearance
of invariant subspaces in $\psik$. We are not interested in such
solution and assume $f_1\ne 0$ for all $N_\zeta$. Equation
\rf{sec5024n2} allows us then to find dependence of the operator
$f_2$ on $N_\zeta$:
\be \label{sec5026} f_2 =
\frac{f_{2(0)}}{(2s+d-2-2N_\zeta)(2s+d-4-2N_\zeta)} \,,\ee
where $f_{2(0)}$ is dimensionfull parameter not depending on
$N_\zeta$. Inserting $f_2$ \rf{sec5026} in \rf{Y2equ} one can make
sure that \rf{Y2equ} is satisfied automatically. All that remains
then is to solve Eq.\rf{Y1equ}. Making use of \rf{msol1}-\rf{deldel3}
and \rf{sec5026} one can make sure that \rf{Y1equ} amounts to the
equation
\footnote{ It is easy to demonstrate that making use of field
redefinitions, the phase factors of $\fwt_3$ can be normalized to be
equal to $-1$. Therefore in \rf{sec5050} and below $\fwt_3$ is
assumed to be real-valued and negative. Relations \rf{DelDelwt} and
Eq.\rf{deldel3} imply then that $\fwt_1$ is real-valued and
positive.}
\beq\label{sec5050}
&& Z(N_\zeta)- Z(N_\zeta-1)
\nonumber\\[5pt]
&& -\frac{(2s+d-3-2N_\zeta) (f_{2(0)})^2 }{(2s+d-2-2N_\zeta)^2
(2s+d-4-2N_\zeta)^2 } - \frac{ \coscon }{4}(2s+ d-3 -2N_\zeta) = 0\,,
\eeq
where we use the notation
\be\label{ZNdef} Z(N_\zeta) \equiv  (2s+d - 4 -2N_\zeta)
(N_\zeta+1)(\fwt_3)^2\,. \ee
The relation \rf{ZNdef} implies a condition $Z(-1)=0$. This condition
and \rf{sec5050} lead to the initial condition
\be \label{Z0} Z(0) = \frac{(2s+d-3) (f_{2(0)})^2 }{(2s+d-2)^2
(2s+d-4)^2 } + \frac{ \coscon }{4}(2s+ d-3 )\,.
\ee Equation \rf{sec5050} and the initial condition \rf{Z0} allow us
to find $Z(N_\zeta)$ uniquely and taking into  account \rf{ZNdef} we
obtain
\be \label{sec5055}
(\fwt_3)^2 =
\frac{2s+d-3-N_\zeta}{2s+d-4-2N_\zeta}\Bigl( \frac{(f_{2(0)})^2 }{
(2s + d -2)^2 (2s+d- 4 - 2N_\zeta)^2} + \frac{\coscon}{4}\Bigr)\,.
\ee
Thus we satisfied all equations imposed on $\MM$ and $\FF$ by the
requirement of gauge invariance of the action and the expressions
\rf{msol1}-\rf{deldel3}, \rf{sec5026}, \rf{sec5055} determine $\MM$
and $\FF$ uniquely. In view of the first relation in \rf{msol1} and
\rf{sec5026} the $f_{2(0)}$ is real-valued and introducing the mass
parameter $\mas$ (which is assumed to be positive) by relation
\be f_{2(0)} = (2s + d - 2)\mas \,,\ee
we arrive at the expressions for $\MM$ and $\FF$ given in Section 2.

To summarize, we found the gauge invariant action for the fermionic
fields in $(A)dS_d$. All that remains to construct action for
fermionic fields in $AdS_5\times S^5$ Ramond-Ramond background is to
add appropriate dependence of $S^5$-coordinates and take into account
contribution of
Ramond-Ramond background fields%
\footnote{ Study of some leading contributions of Ramond-Ramond
background fields to mass operator of the bosonic fields in
$AdS_5\times S^5$ Ramond-Ramond background may be found in
\cite{Burrington:2005ae}. Precise form of mass operator for bosonic
fields is still to understood.}.
The result will be reported elsewhere.

\bigskip
{\bf Acknowledgments}. This work was supported by the INTAS project
03-51-6346, by the RFBR Grant No.05-02-17654, RFBR Grant for Leading
Scientific Schools, Grant No. 1578-2003-2 and Russian Science Support
Foundation.

\setcounter{section}{0} \setcounter{subsection}{0}
\appendix{ Notation and commutators of oscillators and covariant
derivative }

We use $2^{[d/2]}\times 2^{[d/2]}$ Dirac gamma matrices $\gamma^A$ in
$d$-dimensions, $ \{ \gamma^A,\gamma^B\} = 2\eta^{AB}$,
$\gamma^{A\dagger} = \gamma^0 \gamma^A \gamma^0$,
where $\eta^{AB}$ is mostly positive flat metric tensor and flat
vectors indices of the $so(d-1,1)$ algebra take the values
$A,B=0,1,\ldots ,d-1$. To simplify our expressions we drop
$\eta_{AB}$ in scalar products, i.e. we use $X^AY^A \equiv
\eta_{AB}X^A Y^B$. Indices $\mu,\nu=0,1,\ldots d-1$ stand for indices
of space-time base manifold.

We use the algebra of commutators for operators that can be
constructed out the oscillators $\alpha^A$, $\bar\alpha^A$
\rf{intver15} and derivative $D^A$ \rf{vardef01},\rf{lorspiope} (see
also Appendix A in Ref.\cite{Metsaev:1999ui}). Starting with
\be
[\hat{\partial}_A,\hat{\partial}_B]=\Omega_{AB}{}^C\hat{\partial}_C\,,
\qquad \Omega^{ABC}\equiv -\omega^{ABC}+\omega^{BAC}\,, \qquad
\omega_A{}^{BC}\equiv e^\mu_A\omega_\mu^{BC} \,, \ee
where $\hat\partial_A\equiv e_A^\mu\partial_\mu$,
$\partial_\mu\equiv\partial/\partial x^\mu$,  and $\Omega^{ABC}$ is a
contorsion tensor we get the basic commutator
\beq \label{dadbap2}
[D^A , D^B ]  & = & \Omega^{ABC} D^C + \frac{1}{2} R^{ABCD} M^{CD}
\nonumber\\[5pt]
& = & \Omega^{ABC} D^C + \coscon M^{AB}\,,
\eeq
and $R^{ABCD}$ is a Riemann tensor which for $(A)dS_d$ geometry takes
the form
\be R^{ABCD} =  \coscon (\eta^{AC}\eta^{BD} - \eta^{AD}\eta^{BC})\,.
\ee
The spin operator $M^{AB}$ is given in \rf{loralgspiope}. For
flexibility in \rf{dadbap2} and below we present our relations for a
space of arbitrary geometry and for $(A)dS_d$ space. Using
\rf{dadbap2} and the commutators
\be [D^A,\alpha^B]= -\omega^{ABC}\alpha^C\,, \qquad [D^A,
\bar{\alpha}^B]= - \omega^{ABC}\bar{\alpha}^C\,, \qquad [D^A,
\gamma^B]= - \omega^{ABC}\gamma^C\,, \ee
we find straightforwardly
\beq
&& [D^A, \alpha^2]=0\,, \qquad  [\bar\alpha^2, D^A]=0\,,\qquad
[D^A,\gamma\alpha]=0\,,
\\[5pt]
&& [\bar\alpha^2, \alpha D] = 2\bar\alpha D  \,,
\qquad
[\gamma\bar\alpha, \alpha D] = \Dline  \,,
\qquad
\{\Dline, \gamma\alpha\} = 2\alpha D \,,
\eeq
\beq
&& \Dline\!^2 = D^AD^A + \omega^{AAB}D^B +
\frac{1}{4}\gamma^{AB}R^{ABCD} M^{CD}
\nonumber\\[5pt]
&& \hspace{1cm} = D^AD^A + \omega^{AAB}D^B + \frac{ \coscon
}{2}\gamma^{AB}M^{AB}\,,
\\[5pt]
&& [\bar\alpha D , \alpha D ]   = D^AD^A + \omega^{AAB}D^B  -
\frac{1}{4} R^{ABCD} M_b^{AB} M^{CD}
\nonumber\\[5pt]
&& \hspace{1.5cm}  =  D^AD^A + \omega^{AAB}D^B - \frac{\coscon}{2}
M_b^{AB} M^{AB}\,,
\\[5pt]
&& [\Dline,\alpha D ]  = \frac{1}{2}\gamma^A \alpha^B R^{ABCD}M^{CD}
\nonumber \\[5pt]
&&  \hspace{1.5cm} = \coscon \Bigl(  \gamma \alpha (\alpha\bar\alpha
+ \frac{d-1}{2}) - \alpha^2\gamma \bar\alpha\Bigr)\,,
\\[5pt]
&& [\bar\alpha D, \Dline ]  = - \frac{1}{2}\gamma^A \bar\alpha^B
R^{ABCD} M^{CD}
\nonumber \\[5pt]
&&  \hspace{1.5cm} = \coscon \Bigl( (\alpha\bar\alpha +
\frac{d-1}{2}) \gamma\bar \alpha  -  \gamma\alpha\bar\alpha^2 \Bigr)
\,.
\eeq


\newpage
\small
\begin{thebibliography}{30}

\parskip=0.pt


\bibitem{Maldacena:1997re}
  J.~M.~Maldacena,
  Adv.\ Theor.\ Math.\ Phys.\  {\bf 2}, 231 (1998)
  [Int.\ J.\ Theor.\ Phys.\  {\bf 38}, 1113 (1999)].


\bibitem{Metsaev:1998it}
  R.~R.~Metsaev and A.~A.~Tseytlin,
  Nucl.\ Phys.\ B {\bf 533}, 109 (1998)
  [hep-th/9805028].




\bibitem{mt3}
R.~R.~Metsaev and A.~A.~Tseytlin,
Phys.\ Rev.\ D {\bf 63}, 046002 (2001); hep-th/0007036.


\bibitem{Metsaev:1999ui}
  R.~R.~Metsaev,
  Nucl.\ Phys.\ B {\bf 563}, 295 (1999)
  [arXiv:hep-th/9906217].




\bibitem{Metsaev:2003cu}
  R.~R.~Metsaev,
  Phys.\ Lett.\ B {\bf 590}, 95 (2004)
  [arXiv:hep-th/0312297].

\bibitem{Metsaev:2004ee}
  R.~R.~Metsaev,
  Class.\ Quant.\ Grav.\  {\bf 22}, 2777 (2005)
  [arXiv:hep-th/0412311].


\bibitem{Deser:2001xr}
  S.~Deser and A.~Waldron,
  Phys.\ Lett.\ B {\bf 513}, 137 (2001)
  [arXiv:hep-th/0105181].


\bibitem{Fronsdal:1978vb}
  C.~Fronsdal,
  Phys.\ Rev.\ D {\bf 20}, 848 (1979).



\bibitem{Fang:1979hq}
  J.~Fang and C.~Fronsdal,
  Phys.\ Rev.\ D {\bf 22}, 1361 (1980).


\bibitem{Lopatin:1987hz}
  V.~E.~Lopatin and M.~A.~Vasiliev,
  Mod.\ Phys.\ Lett.\ A {\bf 3}, 257 (1988).

\bibitem{Vasiliev:1987tk}
  M.~A.~Vasiliev,
  Nucl.\ Phys.\ B {\bf 301}, 26 (1988).



\bibitem{Metsaev:1995re}
  R.~R.~Metsaev,
  Phys.\ Lett.\ B {\bf 354}, 78 (1995).
%
Phys.\ Lett.\ B {\bf 419}, 49 (1998) [arXiv:hep-th/9802097].


\bibitem{Fotopoulos:2006ci}
  A.~Fotopoulos, K.~L.~Panigrahi and M.~Tsulaia,
  arXiv:hep-th/0607248.



\bibitem{Buchbinder:2001bs}
  I.~L.~Buchbinder, A.~Pashnev and M.~Tsulaia,
  Phys.\ Lett.\ B {\bf 523}, 338 (2001)
  [arXiv:hep-th/0109067].

\bibitem{Biswas:2002nk}
  T.~Biswas and W.~Siegel,
  JHEP {\bf 0207}, 005 (2002)
  [arXiv:hep-th/0203115].


\bibitem{Sagnotti:2003qa}
  A.~Sagnotti and M.~Tsulaia,
  Nucl.\ Phys.\ B {\bf 682}, 83 (2004)
  [arXiv:hep-th/0311257].


\bibitem{Buchbinder:2006ge}
I.~L.~Buchbinder, V.~A.~Krykhtin and P.~M.~Lavrov,
arXiv:hep-th/0608005.





\bibitem{Alkalaev:2003qv}
  K.B.~Alkalaev, O.V.~Shaynkman and M.A.~Vasiliev,
  Nucl.Phys.B {\bf 692}, 363 (2004)
  [hep-th/0311164].

\bibitem{Alkalaev:2005kw}
  K.~B.~Alkalaev, O.~V.~Shaynkman and M.~A.~Vasiliev,
  JHEP {\bf 0508}, 069 (2005)
  [arXiv:hep-th/0501108].

\bibitem{deMedeiros:2003px}
  P.~de Medeiros,
  Class.\ Quant.\ Grav.\  {\bf 21}, 2571 (2004)
  [arXiv:hep-th/0311254].



\bibitem{Hallowell:2005np}
  K.~Hallowell and A.~Waldron,
  Nucl.\ Phys.\ B {\bf 724}, 453 (2005)
  [arXiv:hep-th/0505255].


\bibitem{Gover:2006ha}
  A.~R.~Gover, K.~Hallowell and A.~Waldron,
  arXiv:hep-th/0606160.

\bibitem{Baekler:2006vw}
  P.~Baekler, N.~Boulanger and F.~W.~Hehl,
  arXiv:hep-th/0608122.


\bibitem{Zinoviev:2001dt}
  Yu.~M.~Zinoviev,
  arXiv:hep-th/0108192.

\bibitem{Labastida:1987kw}
  J.~M.~F.~Labastida,
  Nucl.\ Phys.\ B {\bf 322}, 185 (1989).

\bibitem{Buchbinder:2004gp}
  I.L.~Buchbinder, V.A.~Krykhtin and A.Pashnev,
  Nucl.Phys.B {\bf 711}, 367 (2005)
  [arXiv:hep-th/0410215].


\bibitem{Buchbinder:2005ua}
  I.~L.~Buchbinder and V.~A.~Krykhtin,
  Nucl.\ Phys.\ B {\bf 727}, 537 (2005)
  [arXiv:hep-th/0505092].




\bibitem{Francia:2005bu}
  D.~Francia and A.~Sagnotti,
  Phys.\ Lett.\ B {\bf 624}, 93 (2005)
  [arXiv:hep-th/0507144].


\bibitem{Francia:2002aa}
  D.~Francia and A.~Sagnotti,
  Phys.\ Lett.\ B {\bf 543}, 303 (2002)
  [arXiv:hep-th/0207002].

\bibitem{Francia:2002pt}
  D.~Francia and A.~Sagnotti,
  Class.\ Quant.\ Grav.\  {\bf 20}, S473 (2003)
  [arXiv:hep-th/0212185].

\bibitem{Buchbinder:2006nu}
I.~L.~Buchbinder, V.~A.~Krykhtin, L.~L.~Ryskina and H.~Takata,
arXiv:hep-th/0603212.




\bibitem{Deser:1983mm}
  S.~Deser and R.~I.~Nepomechie,
  Annals Phys.\  {\bf 154}, 396 (1984).


\bibitem{Skvortsov:2006at}
  E.~D.~Skvortsov and M.~A.~Vasiliev,
  arXiv:hep-th/0601095.


\bibitem{Deser:2001us}
  S.~Deser and A.~Waldron,
  Nucl.\ Phys.\ B {\bf 607}, 577 (2001)
  [arXiv:hep-th/0103198].



\bibitem{Breitenlohner:1982bm}
P.~Breitenlohner and D.~Z.~Freedman,
Phys.\ Lett.\ B {\bf 115}, 197 (1982).
%
Annals Phys.\  {\bf 144}, 249 (1982).




\bibitem{Deser:2003gw}
S.~Deser and A.~Waldron,
Nucl.\ Phys.\ B {\bf 662}, 379 (2003) [arXiv:hep-th/0301068].
%
Phys.\ Rev.\ Lett.\  {\bf 87}, 031601 (2001) [arXiv:hep-th/0102166].


\bibitem{Burrington:2005ae}
  B.~A.~Burrington and J.~T.~Liu,
  Nucl.\ Phys.\ B {\bf 742}, 230 (2006)
  [arXiv:hep-th/0512151].









\end{thebibliography}
\end{document}